\documentclass[prl,twocolumn,showpacs,amsmath,amssymb,nofootinbib,A4paper]{revtex4}
\usepackage{amsfonts}
\usepackage{amssymb}
\usepackage[dvips]{graphicx}
\usepackage{latexsym}

\newcommand\lsim{\lesssim}
\newcommand\gsim{\gtrsim}

\newcommand\ee{\end{equation}}
\newcommand\be{\begin{equation}}
\newcommand\eea{\end{eqnarray}}
\newcommand\bea{\begin{eqnarray}}

\begin{document}

\title{Strongly Scale-dependent Non-Gaussianity}

\author{Antonio  Riotto$^{1,2}$ and Martin S. Sloth$^{2}$}

\affiliation{\vskip 0.3cm$^1$INFN, Sezione di Padova, Via Marzolo 8,
I-35131 Padua, Italy
\vskip 0.3cm
$^2$ CERN, PH-TH Division, CH-1211, Gen\`eve 23,  Switzerland
}

\begin{abstract}
We discuss models of primordial density perturbations where the non-Gaussianity is strongly scale-dependent. In particular, the non-Gaussianity may have a sharp cut-off and be very suppressed on large cosmological scales, but  sizeable on small scales. This may have an impact on probes of non-Gaussianity in the large-scale structure and in the cosmic microwave background radiation anisotropies.

\end{abstract}
\maketitle
Cosmological inflation 
\cite{lrreview} has become the dominant paradigm to 
understand the initial conditions for the Cosmic Microwave Background (CMB) anisotropies
and Large Scale Structure (LSS) formation. 
This picture has recently received further spectacular confirmation 
by the Wilkinson Microwave 
Anisotropy Probe (WMAP) five year set of data \cite{wmap5}.
Present and future \cite{planck} experiments 
may be sensitive to the non-linearities of the cosmological
perturbations at the level of second- or higher-order perturbation theory.
The detection of these non-linearities through the non-Gaussianity
(NG)  \cite{review} has become one of the primary experimental targets. 

A possible source of NG could be primordial in 
origin, being specific to a particular mechanism for the generation of the cosmological perturbations. This is what 
makes a positive detection of NG so
relevant: it might help in discriminating among competing scenarios which otherwise might be indistinguishable. Indeed,
various models of inflation, firmly rooted in modern 
particle physics theory, predict a significant amount of primordial
NG generated either during or immediately after inflation when the
comoving curvature perturbation becomes constant on super-horizon scales
\cite{review}. While single-field  \cite{noi}
and two(multi)-field \cite{two} models of inflation generically predict a tiny level of NG, 
`curvaton-type models' \cite{Enqvist:2001zp,Lyth:2001nq,Moroi:2001ct},  in which
a  significant contribution to the curvature perturbation is generated after
the end of slow-roll inflation by the perturbation in a field which has
a negligible effect on inflation, may predict a high level of NG \cite{luw}.
Alternatives to the curvaton model are those models 
characterized by the curvature perturbation being 
generated by an inhomogeneity in 
the decay rate \cite{hk,varcoupling} of the inflaton field. 
Other opportunities for generating the curvature perturbation occur
at the end of inflation \cite{endinflation} and during
preheating \cite{preheating}. All these models generate a level of NG which is local, since the NG part of the primordial curvature
perturbation is a local function of the Gaussian part generated on superhorizon scales. 
A phenomenological way of parametrizing the level of NG is to
expand the fully 
 non-linear primordial Bardeen
 gravitational potential, $\Phi$, in powers of the linear gravitational potential,
$\Phi_{\rm L}$,
 \be
 \label{phi}
 \Phi=\Phi_{\rm L}+f_{\rm NL}\left(\Phi_{\rm L}^2-\langle\Phi_{\rm
L}^2\rangle\right).
 \ee
The  dimensionless
quantity  $f_{\rm NL}$  sets the
magnitude of the three-point 
correlation function  \cite{review}.
In momentum space, the three point function (bispectrum), arising from the local NG is dominated by the
so-called ``squeezed'' configuration, where one of the momenta is much smaller than the other two and it is parametrized by
the non-linearity parameter $f_{\rm NL}^{\rm loc}$. Other models, such as DBI inflation
\cite{DBI} and ghost inflation \cite{ghost}, predict a different kind of primordial
NG, called ``equilateral'', because the three-point function for this kind of NG is peaked on equilateral configurations, 
in which the lengths 
of the three wave-vectors forming a triangle in Fourier
space are equal \cite{Shapes}. The equilateral NG is parametrized by an amplitude $f_{\rm NL}^{\rm equil}$~\cite{CN}. Present limits
on NG are summarized by $-4<f^{\rm loc}_{\rm NL}<90$ and   $-151<f^{\rm equil}_{\rm NL}<253$ at 95\% CL \cite{wmap5,Curto,zal}. 

It is clear that  detecting a
significant amount of 
NG and its shape either from the CMB or from the
LSS offers the possibility of opening a window into the 
dynamics of the universe
during the very first 
stages of its
evolution and to understand what mechanism gave rise to the cosmological perturbations.
Besides in the CMB anisotropies, where the Signal-to-Noise Ratio (SNR) for the NG is dominated by
the small (angular) scales \cite{review,rec}, the primordial NG
alters the clustering of dark matter halos inducing a scale-dependent
bias on large 
scales \cite{Dalal}.  Furthermore, non-Gaussianities are particularly relevant in the high-mass end of
the power spectrum of perturbations, i.e. on the scale of galaxy clusters,
since the effect of NG fluctuations becomes especially visible on
the  tail of the probability distribution \cite{tail}. 
As a result, both the clustering properties and abundance of very massive halos
are sensitive probes of primordial 
non-Gaussianities 
and could be detected or significantly constrained by
the various planned large-scale galaxy surveys,
both ground based (such as DES, PanSTARRS and LSST) and in space
(such as EUCLID and ADEPT) \cite{komatsuprop}.

The fact that the CMB anisotropies and the LSS probe NG at different cosmological scales has 
recently spurred quite a lot of interest in the possibility that NG is running with the cosmological scale \cite{running1,running2,running3}.
For instance, single-field models with changing sound speed can have strongly scale-dependent NG \cite{running1}. Such models could evade the CMB constraints, but still have important effects at scales responsible for the formation of cosmological objects such as clusters and galaxies. A simple power law ansatz to describe the scale-dependence of NG is asumed for the bispectrum
$\langle\Phi(k_1)\Phi(k_2)\Phi(k_3)\rangle=f_{\rm NL}\left(K/k_{\rm p}\right)^{n_{f_{\rm NL}}}F(k_1,k_2,k_3)$, where 
$K=(k_1 k_2 k_3)^{1/3}$, $k_{\rm p}$ is some fixed pivot scale and $n_{f_{\rm NL}}$ is the NG spectral index.
Future observations such as CMBpol may be able to detect the scale-dependence of NG. The observational error,  around the fiducial value $n_{f_{\rm NL}}=0$, is $\Delta n_{f_{\rm NL}}\simeq 0.05
(50/f_{\rm NL})/\sqrt{f_{\rm sky}}$ \cite{running2}, where $f_{\rm sky}$ is the fraction of the observed sky.
The reason why the power-law ansatz has been adopted  in the literature is that the 
scale-dependence of NG turns out to be moderate in all models analyzed, $\left|n_{f_{\rm NL}}\right|\lsim 1$, and usually proportional to the slow-roll parameters \cite{running3}.

The goal of this short note is to show that the scale-dependence of the NG can naturally be much stronger than what have been discussed in the literature so far. 
The NG can be sizeable at small cosmological scales and totally negligible at large scales 
with the transition between the two regimes taking place at some comoving scale $\sim k_*^{-1}$ exiting the Hubble radius
during the last 60 (or so) e-folds of inflation. If one insists on making use of the NG spectral index, such a situation might be described by   
$n_{f_{\rm NL}}\simeq 0$ for comoving momenta $k\gsim k_*$ and $n_{f_{\rm NL}}$ large and positive for momenta $k\lsim k_*$. The reason why one is interested in such a possibility is readily explained.
The square of the SNR corresponding to a non-vanishing NG bispectrum in the CMB anisotropies is proportional to
\be
\int d^2\ell_1d^2\ell_2 d^2\ell_3\delta^{(2)}(\vec{\ell}_1+\vec{\ell}_2+\vec{\ell}_3)\frac{B^2(\ell_1,\ell_2,\ell_3)}{C(\ell_1)C(\ell_2)C(\ell_2)},
\ee
where, for simplicity, we have assumed the flat-sky approximation, $C(\ell)$ and $B(\ell_1,\ell_2,\ell_3)$ are the amplitudes of the two- and three-point temperature anisotropy correlators in multipole space, 
respectively. Consider,  for example, the case in which the NG is dominated by the squeezed configuration (local case), $\ell_1\ll \ell_{2,3}$. For $\ell\gg\ell_{\rm D}\simeq 750$ the combination of the radiation driving and Silk damping causes the  exponential  decrease of the angular power spectrum and bispectrum such that  square of the SNR turns out to be proportional to \cite{rec} 
\be
(f_{\rm NL}^{\rm loc})^2\int \frac{d^2\ell_1}{\ell_1^3}d^2\ell_2 \sim (f_{\rm NL}^{\rm loc})^2\left(\frac{\ell_{\rm D}}{\ell_{\rm min}}\right)\,\ell^2_{\rm max},
\ee
where $\ell_{\rm min}$ is a multipole of the order of $ \ell_{\rm D}$ and $\ell_{\rm max}$ is the maximum multipole a given experiment can reach. Therefore, the SNR is dominated by the smallest
(angular) scale. The same is true for other shapes of NG; for instance for the equilateral NG, where
the NG is a maximized for $\ell_1\sim\ell_2\sim\ell_3$, the square of the SNR scales like
$(f_{\rm NL}^{\rm eq})^2\ell_{\rm max}$. On the LSS side, the primordial local NG
manifests itself by altering the clustering of dark matter peaks (halos) and inducing a scale-dependent Lagrangian
bias between the power spectrum of halos and dark matter, $b^2_{\rm L}=P_{\rm h}(k)/P_{\rm dm}(k)$ on large scales \cite{Dalal}. The forecast from the shape of the large-scale
power spectrum in the presence of NG in a given galaxy survey is 
sensitive to the largest scale  $\sim k_{\rm min}^{-1}$ within a given redshift slice and the 
square of the SNR scales like $(f_{\rm NL}^{\rm loc})^2/k^4_{\rm min}$ \cite{car}.
The LSS bias probe is sensitive to the large scale side of the primordial NG as the long
wavelength part of the second-order perturbation modulates the position of the dark matter density peaks where galaxy reside. Therefore, a model
of local primordial NG where the non-linearities are switched on at  small cosmological scales and off at large scales will give no signal
in the LSS bias,  while it will possibly manifest itself in the CMB anisotropy.

Let us now see how  such a situation may be easily realized. Consider a  scalar field $\sigma$ which gives a subdominant contribution to the energy density and which is initially heavy with mass $M \gg H$, but which  becomes light,  $m\ll H$ at some (conformal) time $\tau=\tau_*$, corresponding to the time when the comoving wavelength  $1/k_*$ is exiting the Hubble radius. Here $H$ indicates the approximately constant value of the Hubble rate during inflation. 
We will assume that the linear perturbations are dominated by the inflaton fluctuations, and since the $\sigma$-field gives a sub-dominant contribution to the total energy density, the comoving curvature perturbation after the inflaton has decayed into radiation is $
\zeta \approx  \zeta_{\rm rad} + \delta\rho_\sigma/(4 \rho_{\rm tot})$, 
where we used $\zeta\equiv -H\delta\rho/\dot\rho$ in spatially flat gauge, and that after inflation we have $\dot\rho_{\rm tot} = -4 H\rho_{\rm tot}$. 
Since the perturbation in the radiation fluid seeded by the inflaton is approximately conserved, and can be computed the usual way during inflation, we only need to compute $\delta\rho_\sigma$ after the end of inflation in the two different cases: first for modes crossing outside the horizon when the mass of the $\sigma$-field is $M\gg H$;  secondly for modes crossing outside the horizon when the mass of the new field is $m\ll H$.

Let us first consider the modes which exited the horizon with mass $M\gg H$. The solution of the mode equation
during inflation is
\be
\sigma_{\bf k} = -H\tau^{3/2}\left[C_1 H_{\nu_M}^{(1)}(-k\tau)+C_2H_{\nu_M}^{(2)}(-k\tau)\right],
\ee
where $\nu_M^2 = 9/4-M^2/H^ 2$ (so that we have $\nu_M\sim iM/H$ for $M\gg H$) and 
$H_{\nu_M}^{(1,2)}(x)$ are the Hankel functions. 
Picking up the Bunch-Davies vacuum for  modes  well inside the Hubble radius gives
$C_1 = (\pi/2) {\rm exp}\left(-\pi M/2H\right)$ and  $ C_2=0$. 
At late times, when  $(-k\tau)\ll1$, the Hankel function behaves as 
\be
\left| H_{i \frac{M}{H}}^{(1)}(x\ll 1) \right|^ 2 \sim \frac{2}{\pi}\frac{H}{M}e^{\pi M/H},
\ee
and therefore the solution for $(-k\tau_*)\ll 1 $ is
\be
\left|\sigma_{\bf k}(\tau\ll\tau_*)\right|^2\approx \frac{\pi}{2}\frac{H}{M}\frac{H^ 2}{k^ 3}\left(\frac{k}{aH}\right)^ 3.
\ee
After the mass becomes small the field almost freezes, and we have for super-Hubble scales at the end of inflation $\tau_{\rm e}\gg\tau_*$
\be \label{sol0}
\left|\sigma_{\bf k}(\tau_{\rm e})\right|^2\approx \frac{1}{2}\frac{H^3}{Mk^3}\left(\frac{k}{a_*H}\right)^ 3\left(\frac{a_*H}{a_{\rm e}H}\right)^{3-2\nu_m},
\ee
where $\nu_m \sim  3/2-m^2/(3H^2) $ and we will for simplicity assume that $m^2/(3H^2) \gg \epsilon\equiv -\dot H/H^2$. 
On the other hand, modes that exit the horizon after $\tau_*$ with  $(-k\tau_*)\gg 1$ freeze in the moment they exit the Hubble radius in the usual way, and at the end of inflation on super-horizon scales they have the usual almost scale invariant spectrum given by
\be\label{sol1}
|\sigma_{\bf k}(\tau_{\rm e})|^2 \approx \frac{H^2}{2 k^{3}}\left(\frac{k}{a_{\rm e}H}\right)^{3-2\nu_m}.
\ee
After the end of inflation when the inflaton has decayed into radiation, the Hubble rate will rapidly decrease, becoming smaller than $m$, and the $\sigma$-field will quickly become non-relativistic. At this stage the amplitude $|\sigma_{\bf k}|^2$ will again begin to decrease like $1/a^3$. Thus, during radiation domination, for $\tau\gg\tau_{\rm e}$ , we can obtain the respective solutions for $|\sigma_{\bf k}|^2$ averaged over an oscillation period by multiplying Eqs. (\ref{sol0}) and (\ref{sol1}) by a factor $(1/2)(a_{\rm e}/a)^3$.

The vacuum expectation value of the $\sigma$-field will very quickly be driven to zero during the first stage of inflation when the mass of the field is large. Thus, the energy density fluctuations of the $\sigma$  field becomes 
\be
\frac{\delta\rho_\sigma(k)}{\rho_{\rm tot}} = \frac{\frac{1}{2} m^ 2 \left|\sigma_{\bf k}\right|^2}{\rho_{\rm tot}},
\ee
Note that the modes for which $k\ll a_*H$  are suppressed by a factor $H/M (k/a_* H)^ 3$ compared those for which  $k\gg a_* H$. However, even the modes that exit when the mass is small $k\gg a_* H$  give a small contribution to the curvature perturbation at the end of inflation, but after the end of inflation they decrease slower than the background energy density of radiation and regain relative strength. Indeed, 
defining the power spectrum $
\langle\sigma^2_{\bf k} \sigma^2_{\bf q}\rangle \equiv (2\pi)^3\delta^{(3)}({\bf{k}}+{\bf{q}})\frac{2\pi^2}{k^3}{{\mathcal{P}}}_{\sigma^2}(k)$  
and setting $\eta_{\sigma}=2m^2/(3 H^2)$, 
we can write
\be
\mathcal{P}_{\zeta}(k) = \mathcal{P}_{\zeta_{\rm rad}}(k) +\frac{1}{256}\eta_{\sigma}^2\left(\frac{a}{a_{\rm e}}\right)^8\frac{1}{M_p^4}{\mathcal{P}}_{\sigma^2}(k)
\ee  
and following a calculation similar to that of Ref.  \cite{Boubekeur:2005fj}, one can show that 
\be
{{\mathcal{P}}}_{\sigma^2}(k)\simeq \frac{k^3}{2\pi}\left(\frac{H}{2\pi}\right)^4\frac{1}{4}\left(\frac{a_{\rm e}}{a}\right)^6\int\frac{d^3p}{(p~|{\bf p}-{\bf k}|)^{3-\eta_\sigma}},
\ee
which implies  ${{\mathcal{P}}}_{\sigma^2}(k)\simeq (6/4\eta_\sigma)(a_{\rm e}/a)^6{{\mathcal{P}}}_{\sigma}^2(k)$.

The bispectrum of the curvature perturbation can be defined by
\be
\left<\zeta_{k_1}\zeta_{k_2}\zeta_{k_3}\right> = (2\pi)^3\delta^{(3)}({\bf{k}}_1+{\bf{k}}_2+{\bf{k}}_3)B_{\zeta}(k_1,k_2,k_3)
\ee
such that a local type NG can be parametrized by a $f^{\rm loc}_{\rm NL}$ parameter by
\be
B_{\zeta}(k_1,k_2,k_3) =\frac{6}{5}f^{\rm loc}_{\rm NL}\left[P_{\zeta}(k_1)P_{\zeta}(k_2)+\textrm{cyclic perm.}\right],
\ee
where $ P_\zeta\equiv (2\pi^2/k^3){\mathcal P}_\zeta$.  
We will assume that the NG from the inflaton is negligible and that the fluctuations of the inflaton are  uncorrelated with those of the $\sigma$-field. Under these assumptions, the bispectrum of the curvature perturbation is given by
\be
B_{\zeta}(k_1,k_2,k_3) \simeq       \frac{1}{4096}\eta_{\sigma}^3\left(\frac{a}{a_{\rm e}}\right)^{12} \frac{1}{M_p^6} B_{\sigma^2}(k_1,k_2,k_3),
\ee
where 
\be
B_{\sigma^2}=\frac{1}{(2\pi)^ 3}H^6\frac{1}{8}\left(\frac{a_{\rm e}}{a}\right)^{9}
\int  \frac{d^3 p}{(p~ |{\bf p}-{\bf k}_1|~|{\bf p}-{\bf k}_2|)^{3-\eta_{\sigma}}}.
\ee
In the squeezed limit $k_3 \ll k_1\simeq k_2\simeq k$, it reduces to   
\be
B_{\sigma^2}\simeq 
\frac{5\pi}{8}\frac{1}{\eta_{\sigma}}\frac{(2\pi)^3}{k^6}\left(\frac{H}{2\pi}\right)^6\frac{1}{8}\left(\frac{a_{\rm e}}{a}\right)^{9}
\ee
which implies 
\be
f^{\rm loc}_{\rm NL} \approx 2.5 \times10^{-5}(2\epsilon)^3\eta_{\sigma}^2 \mathcal{P}_{\zeta}\left(\frac{a}{a_{\rm e}}\right)^3. 
\ee
The contribution to the curvature perturbation will be subdominant when the $\sigma$-field decays if
$
(1/4)(6/256)(2\epsilon)^ 2(a_{\rm d}/a_{\rm e})^2\eta_\sigma \mathcal{P}^2_{\zeta}\ll\mathcal{P}_{\zeta}$, with the scale factor at the time of decay denoted by $a_{\rm d}$. Therefore, 
the $\sigma$-field  maximally contributes  to the spectrum say at the level of $\sim{\cal O}(10)$\%,  without disturbing the spectrum, if  $
(a_{\rm e}/a_{\rm d}) \gtrsim  \epsilon \eta^{1/2}_\sigma\mathcal{P}^{1/2}_{\zeta}$. 
We then obtain that the NG is suppressed at large scales and is in the observationally interesting range
\be
\left| f^{\rm loc}_{\rm NL}\right| \lsim  2\times10^{-3}\eta_{\sigma}^{1/2} \mathcal{P}^{-1/2}_{\zeta}={\cal O}(10^2) 
\ee
for  comoving  wavenumbers $k\gsim k_*$.   This simple set-up therefore predicts
a NG which is basically vanishing at large scales and switched on at small cosmological scales. 
Let us now come back to the issue of how we can engineer a model where the mass of the $\sigma$-field changes abruptly at some time $\tau_*$ within the last 60 e-folds or so. Suppose that, beyond the
inflaton $\phi$ and the $\sigma$-field, there is a third field, $\chi$, such that the full potential is
\be
V=V(\phi)+\frac{1}{2}\left(g^2\chi^2+m^2\right)\sigma^2+\frac{1}{2}\left(-m_\chi^2+h^2\phi^2\right)\chi^2+\frac{1}{4}\lambda\chi^4.
\ee
If inflation is of the small-field type \cite{lrreview} (it may be of the hybrid type or not), then in the first stage of inflation
the $\chi$ field acquires a vacuum expectation value $\langle\chi\rangle^2=m_\chi^2/\lambda$, and the effective mass squared of the $\sigma$-field becomes $M^2\sim  (g^2 m_\chi^2/\lambda)$. It will be larger than $H^2$ if $(g^2/\lambda)\gsim (H/m_\chi)^2$. As soon as $\langle \phi\rangle^2$ becomes larger than $m_\chi^2/h^2$ the vacuum expectation value of the $\chi$ field vanishes and the mass squared of the $\sigma$ becomes $m^2$ which is assumed to be smaller than $H^2$ during inflation. The wavenumber $k_*$ 
at which the NG are switched on corresponds to the scale exiting the Hubble radius when $\langle \phi\rangle^2\sim m_\chi^2/h^2$. Notice that the same pattern can be obtained in large-field models of inflation (again,  it might be of the hybrid type), it suffices to have $m^2_\chi<0$ and to change the sign of the $\phi^2\chi^2$ interaction. 

Let us close with a couple of comments. First, the transition from the massive to the massless regime
needs a fine-tuning to happen during the last 60 e-folds or so; 
nevertheless, if inflation is of the hybrid type, taking the parameters in the same ballpark of those
added here, will naturally lead to a transition within the observationally interesting range
of scales. 
Secondly, we have found  it hard to
build up models where the transition is from massless to massive regime, corresponding to
a sizeable NG at large scales and suppressed at small scales: the NG perturbations generated on large scales tend to be overwhelmingly diluted in the subsequent 
evolution. Of course, one can turn this apparently negative comment upside down and conclude
that strongly scale dependent NG will most preferably show up in the CMB.

{\it Acknowledgments.}~ 
A.R. acknowledges partial support by the EU Marie Curie Network UniverseNet (HPRNCT2006035863).

\end{document}